\documentclass[12pt]{article} 
\input epsf.tex 
\def\be{\begin{equation}} \def\ee{\end{equation}}
\def\bi{\begin{itemize}} \def\ei{\end{itemize}}
\def\bea{\begin{eqnarray}} \def\eea{\end{eqnarray}} \def\ba{\begin{array}}
\def\ea{\end{array}} \def\ben{\begin{enumerate}} \def\een{\end{enumerate}}

\newcommand{\eqn}[1]{(\ref{#1})}

\newcommand{\prl}[3]{Phys. Rev. Lett. {\bf#1} ({#2}) {#3}}

\newcommand{\hepth}[1]{{\tt arXiv:{#1}[hep-th]}}
\newcommand{\arxiv}[1]{{\tt arXiv:{#1}[hep-th]}}

\def\ep{\epsilon}

\def\br{\nonumber\\}

\begin{document}
{}~
\hfill \vbox{
\hbox{arXiv:2210.13970} 
\hbox{25/10/2022}}
\break

\vskip 3.5cm
\centerline{\large \bf
Islands and Icebergs may contribute nothing to the Page curve}

\vskip 1cm

\vspace*{1cm}

\centerline{\sc  Harvendra Singh }

\vspace*{.5cm}
\centerline{ \it  Theory Division, Saha Institute of Nuclear Physics (HBNI)} 
\centerline{ \it  1/AF Bidhannagar, Kolkata 700064, India}
\vspace*{.25cm}

\vspace*{.5cm}

\vskip.5cm


\centerline{\bf Abstract} \bigskip

We study the entanglement entropy of a subsystem in contact with 
symmetrical bath where the complete system lives on the boundary of
AdS3 spacetime. The system-A is taken to be in the middle of 
the bath system-B and the full system is taken to be some fixed
localized region of the boundary 2-dimensional CFT. 
We generally assume that the
d.o.f.s in the total system remain fixed when we vary the 
size of the bath which is to be guided by the conservation laws. 
It is found that the island and the subleading (icebergs) contributions are 
inseparable, and in totality they 
contribute nothing to the Page-curve of the bath. 
As such they contribute only to the unphysical branch of the entropy. 
The quantum entropy formula 
of the bath may simply be written as ${}_{min}\{S[A],S[A]+Const\}$, 
including for the black holes.
 
\vfill 
\eject

\baselineskip=16.2pt


\section{Introduction}

 The holographic principle in string theory  \cite{malda} has 
revolutionised our understanding of strongly coupled quantum field theories. 
We shall mainly focus on the phenomenon of entanglement between 
two similar looking quantum systems having a common interfaces. 
The quantum information sharing  is 
a real time phenomenon as the subsystem states always remain entangled.
The quantum mechanical states evolve over the time. 
In quantum systems the information  contained in a given state
cannot be destroyed, cloned or even mutated. For example, in simple
bi-partite systems the information can either be found 
in one part of the Hilbert space  or in the compliment \cite{pati,pati1}.
Generally  these exchanges or sharings of 
quantum information is guided by the unitarity and locality. 
Under such claims the understanding of the 
formation of gravitational black holes and subsequent
evaporation processes (via Hawking radiation) remains a long puzzle. 
It is generally believed that the whole process should still be unitary and all
information can be recovered after the black hole has fully evaporated. 
There is a proposal that the entanglement entropy curve for the
radiation should bend after the half Page-time is crossed \cite{page}. 
This  certainly  holds good when a pure state is divided into two smaller subsystems. 
But for  mixed states, or finite temperature CFT duals of the AdS-black holes,
it is not so straight forward to answer this question. 
However, an important progress has been made in some recent models by coupling 
holographic CFT to an external bath (or radiation) system, and also by
involving  nonperturbative  techniques such as 
replica, wormholes and islands \cite{almheri,replica19}. Some
answers to these difficult questions have been attempted.
\footnote{ Also see a review on information paradox along
different paradigms in \cite{raju}
and for a list of related references therein; 
see also  [\cite{susski}-\cite{hashi}].} 

Particularly the recent proposal for generalised entanglement
entropy \cite{almheri} involves an hypothesis of the island $(I)$ 
contribution, including the gravitational
 contribution from the island boundary 
$(\partial I)$, 
such that the complete quantum entropy of radiation bath $(B)$ 
can be expressed as
\bea\label{ficti1}
 S_{Rad}[B]=_{min}[ ext\{ {Area(\partial I)\over 4G_N} + S[B ~U ~I]\}] 
\eea
Thus it is a `hybrid model' as it involves contribution
of gravitational `islands' in dual JT gravity. There are both field 
theoretic and gravitational contributions in it. The formula however
 suggests that one  needs to pick the lowest contribution 
out of a set of many such possible extremas, which inevitably
 includes island entropy contributions.\footnote{In some extensions of the
hybrid models one might also include wormhole contributions, see \cite{replica19}.} 
The complicated looking 
formula \eqn{ficti1} seemingly reproduces a Page-curve for the radiation entropy 
\cite{almheri}. But it ignores contributions of the infinitely many such
sub-leading terms which we shall pronounce here together as '{\it icebergs}' terms. 
One of the important feature of above proposal entails in the 
appearance of islands inside the dual (JT) gravity, 
usually outside of black hole horizon. Although 
the island itself does not arise by means of a dynamical principle. 
These are supposed to be associated with 
the presence of a bath system. The problem we highlight is that
other similar subleading contributions of icebergs have been neglected. 
In this work we present an 
alternative view that there would exist island term but we must
not ignore important icebergs contributions to the entropy. 
If these are ignored 
we may end up with an incorrect picture of the Page curve. 

We are able to systematically 
show that the island and icebergs alongwith the leading entropy term
altogether add up nicely as a series to give just a constant 
contribution to the entropy of the bath. 
These contributions are thus naturally inseparable 
from each other and also they compensate 
each other perfectly no matter what their individual contributions might be.  
 We explicitly show this for the limiting case of a quantum-dot  
at the interface of symmetrical CFT (bath) system. 
 Correspondingly  there would be 
geometric contributions to the total entropy
arising out of island and the icebergs, for a subsystem
with large bath, such that total entropy of both systems is
$$
S_{total}[A U B]=S^{(0)}_{bath}+S_{Island}+S_{icebergs}={\rm Fixed}
$$
Furthermore, the quantum formula for the entropy of bath subsystem
may simply be written  as
$$ S_{quantum}[B]=_{min}\{S[A], S[B]\}   $$
The above expressions reproduce the Page curve for bath, including
for the finite temperature case. 
The formula is valid for the static (equilibrium) situations. 
In explicit time dependent cases, like
black hole evaporation, if at all there is change in the 
total system size (i.e. due to net loss or gain of d.o.fs) 
the picture would be similar at any given instant of time,
so it may be applicable for slow processes only.

The paper is organized as follows. In section-2 we introduce 
the new icebergs contributions and define the generalized entropy
formulation for pure $AdS_3$ case. On the boundary we take
a finite subsystem in contact with a symmetrical bath.
We then discuss a limiting case when subsystem becomes a dot like
and in contact with finite size symmetrical CFT bath. The
situation emerges like that of $n$ quantum dots and a bath.
We extend our results for the black holes case in section-3. 
The last section-4 contains a brief summary.   

\section{The Islands and Icebergs?}

Let us take a subsystem (A) in contact 
with a symmetrical  bath system (B), both of finite size, 
living on the 
boundary of $AdS_3$ spacetime. It is important that both systems are made of 
identical species, i.e. have identical field content, for simplicity. 
Consider pure $AdS_3$ spacetime geometry
\bea\label{ads3}
ds^2={L^2 \over z^2} (- dt^2 + dx^2 + dz^2)
\eea
where $L$ is large  radius of curvature. 
The coordinate ranges are $-\infty\le (t,~x)\le\infty$ 
and  $0\le z \le \infty$ 
represents the full holographic range.\footnote{ The Kaluza-Klein
compactification on a circle ($x\simeq x+ 2\pi R$)
 produces a
{\it nearly or rather conformally } $AdS_2$  solution,  
also well known as Jackiw-Teitelboim 
 background \cite{JT,JT1}, which we note down here
\bea\label{gy67}
&& ds^2_{JT}={L^2 \over z^2} (- dt^2 + dz^2)\br
&& e^{-2 (\phi-\phi_0)}= \sqrt{g_{xx}}={L \over z} 
\eea
where $\phi$ is the $2d$ dilaton field of JT theory, 
written in standard convention 
(effective string coupling vanishes near the boundary). The two
 Newton's constants
get related as ${2\pi R \over G_3}\equiv {1 \over G_2}$, 
with $G_2$ being dimensionless in 2-dim.
 The anti-de Sitter solution without a dilaton 
 remains topological spacetime in 2d with no propagating degrees of freedom. }
\begin{figure}[h]
\centerline{\epsfxsize=3.2in
\epsffile{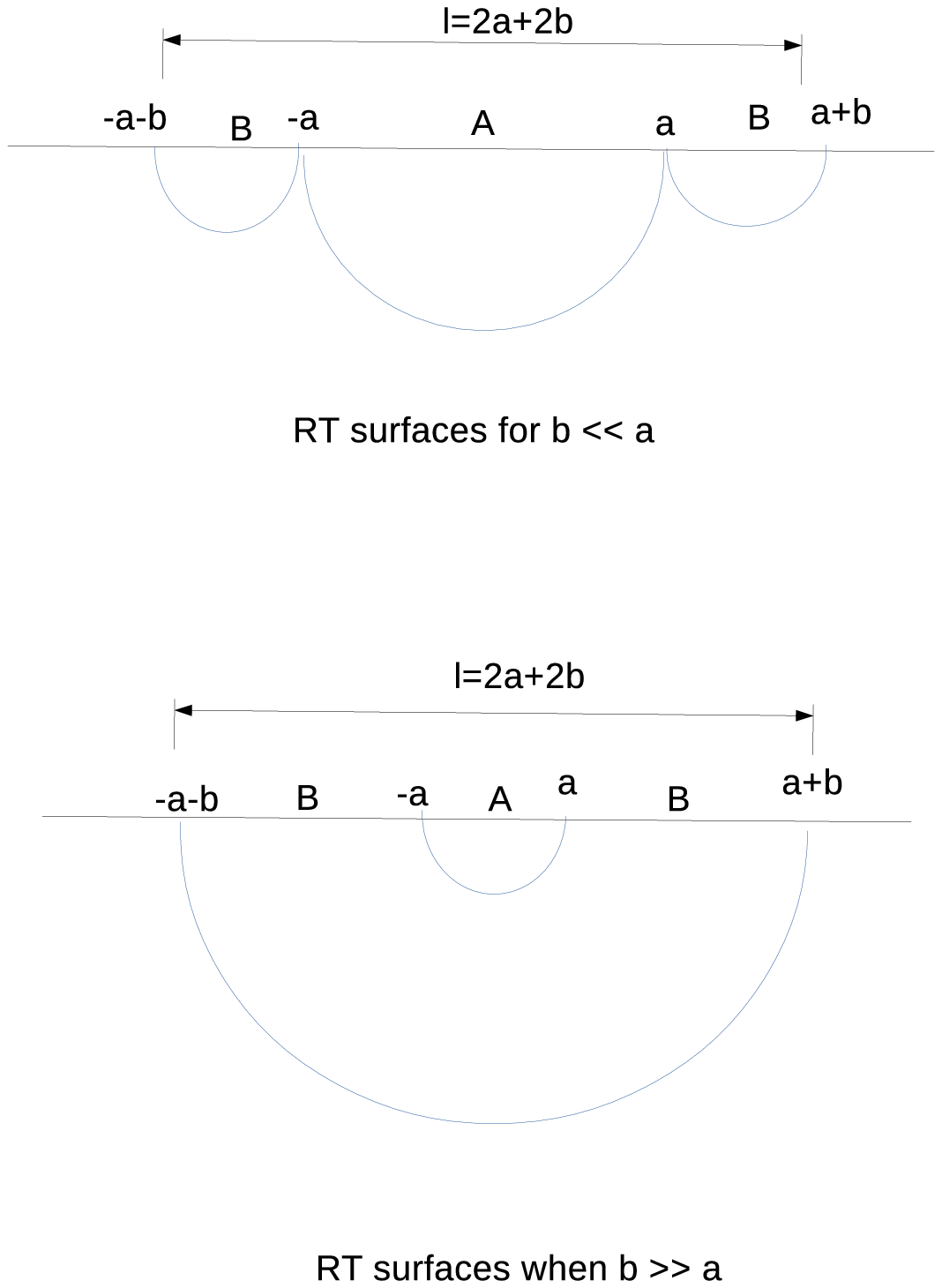} }
\caption{\label{fig22b} 
\it 
Two extremal situations are drawn separately. The extremal RT surfaces when bath size
is much small compared to the subsystem-A ($b\ll a$), we have 
disconnected minimal area surfaces.
The lower graph (for $b\gg a$) instead has connecting surfaces. 
Note we should keep $l (=2a+2b)$  same for the both cases. 
}
\end{figure} 
 
The $CFT_2$ lives on entire 2-dimensional noncompact 
$(t,x)$ flat boundary of $AdS_3$.
The CFT bath system lives on the coordinate patches
$[-(b+a),-a]$  $[a, (b+a)]$ 
along spatial $x$ direction, with
subsystem-A sandwiched in between $[-a,a]$; 
see the sketches in figure \eqn{fig22b}. 
The entire system setup is taken in a particular symmetrical way
for the convenience. The states of the system-A and the bath 
are necessarily entangled. 

It is clear that the entanglement entropy of an extremal $CFT_2$  system
 of net size $l(=2a+2b)$ is given by 
\bea\label{tot1}
S_{total}[AUB]= {L \over 2 G_3}\ln{2(a+b) \over \epsilon}
\eea
where $\epsilon$ represents the UV cut-off scale. 
We shall keep $l$ sufficiently large and fixed  through out, 
and would like to only vary $b$ between $0$ and $l/2$. 
We, of course, assume any local exchanges, that is 
gain (loss) of d.o.f. (information) 
in system-A is compensated by equal loss (gain) in the size of bath 
and vice versa. For dynamical cases 
there might be some definite rate of change, but $\dot a=- 2 \dot b$
would be true due to local conservation laws also.
The exact rate of loss or gain and the mechanism 
is not important and the actual nature of 
the physical process is not required here. 
All we are considering is that
local conservation laws are at work, for the entire system $l$.
We are not considering explicit time dependent processes here. 
Obviously we are assuming that 
subsystem and bath are made of identical (CFT) fields content.
The complementary system includes the patches 
$[b+a,\infty]$ and $[-(b+a),-\infty]$. 

We will only study the static situations.
Consider now two extreme type of cases below.
 
{\it Case-1:} When $b\ll a$, i.e. the bath size is very small, 
the HEE of the system-A is given by respective extremal RT surface \cite{RT,HRT}
\bea
S[A]= {L \over 2 G_3}\ln{(l-2b) \over \epsilon}
\eea
while that of full bath system-B (on both sides) becomes
\bea\label{fin1}
S[B]= {L \over  G_3}\ln{b \over \epsilon}
\eea
 This expression is the standard bath entropy  and no
other extremization is needed. The entropy of the bath will
follow the equation \eqn{fin1} as $b$ grows large. 
However, 
at some point for  large size bath a different extremum
situation emerges, as the new entremal surfaces will appear.

{\it Case-2:} For $b\gg a$ situation, one finds that the bath subsystem-B 
the entanglement entropy instead becomes
\bea\label{fin2}
S[B]= 
{L \over 2 G_3}\ln{l \over \epsilon}
+{L \over 2 G_3}\ln{l-2b \over \epsilon}
\eea
as there is now an RT surface connecting the two farther ends 
of the bath system. The first term in \eqn{fin2} 
is in fact the entropy $S[AUB]$ of the full system `subsystem-A and the bath-B'. 
However $S[A U B]$ is independent of the individual sizes
 ($b$ or $a$), and  it
is  a fixed quantity also for given $l$.
While the subsystem-A entropy is given by, 
\bea\label{fin2k}
S[A]={L \over 2 G_3}\ln{l-2b \over \epsilon}
\eea
note we can write $(a={l\over 2}-b)$ whenever required.
Now since $l$ is fixed from beginning, both $S[A]$ and $S[B]$ 
above qualify as two independent extremas for bath subsystem-B entropy. 
Note it is because
these entropies in \eqn{fin2} and \eqn{fin2k} have same $b$-dependences.
However it is clear that $S[B]>S[A]$, therefore the HEE of the
 large bath  should be  given by the following quantum minimality principle 
\bea\label{fin2s}
S_{quantum}[B]={}_{min}\{S[A],S[B]\}=S[A] \ . \eea 
So the  Page curve  for the bath subsystem-B
follows from the
principle that the minimum entropy, if there exist many
 possible extremas having identical local dependences, will only be accepted. 
This is the main conclusion of eq.\eqn{fin2s}. 

This forms  complete result for extremal AdS case  (i.e. zero temperature 
CFT) of  quantum dot system in contact with symmetrical bath. 

 \begin{figure}[h]
\centerline{\epsfxsize=3in
\epsffile{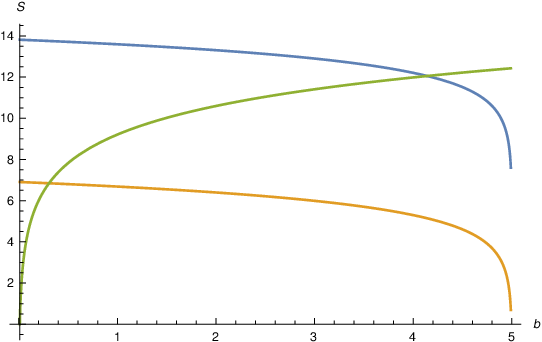} }
\caption{\label{fig5aa}
\small\it
Plots of $Log[l/\ep] + Log[(l - 2 b)/\ep]$, $Log[(l - 2 b)/\ep]$, and 
$2 Log[b/\ep]$
for the values $l=10,~\ep=.01$, for  2-dim CFT system. The
falling curve (lower yellow) is preferred for entanglement entropy
of a larger bath $(b\gg 0)$ under quantum entropy proposal. 
The sole rising curve (green) is good for entropy of small size bath
system only. We see that for large size baths entropy falls towards the end.
The  topmost graph valid in large bath region is  unphysical. Two falling curves
differ by an overall constant only in large $b$ region.
We set $ {L\over 2 G_3}=1$.}
\end{figure}

\noindent{\bf Finding Island and the Icebergs?}

Note that we have not encountered any island or icebergs
so far. Where are these contributions hidden in the above analysis? We 
have got the entropy Page curve  without even knowing  these individual
quantities.
So let us discuss the total entropy of the `subsystem and its bath'
which we have denoted by $ S[A U B]$.
Making an expansion
of the r.h.s of \eqn{tot1}, in  small ratio $({a\over b})$,
 for $a \ll b$, one can find that
\bea\label{fin2ab}
S[A U B]
&= &
{L \over 2 G_3}\left( \ln{2b \over \epsilon} + {a\over  b}-\underbrace{ 
{1\over 2} ({a\over b})^2 +\cdots }_{Iceberg~terms}\right)\br
&\equiv &
S^{(0)}_{bath}
+ S_{Island}
+ S_{Icebergs}
\eea
where expressions in last equality can be identified as
\bea
&&S^{(0)}_{bath}={L \over 2 G_3} 
\ln{2b \over \epsilon},~~~ ~~S_{Island}=
{L \over 2 G_3} {a\over b}, ~~~
\br &&S_{Icebergs}=-
{L \over 4 G_3} \big[ ({a \over b})^2-{2\over3} ({a \over b})^3  + \cdots \big]. 
\eea
Note the very first term on r.h.s. of eq.\eqn{fin2ab} represents
purely the bath system entropy of a box size $2b$,
while  subleading term $S_{Island}$ represents the interaction between 
bath and the subsystem and it may be recognized as twice of 
`gravitational' entropy due to an `Island' boundary (that is 
located at $z=b$ inside the AdS bulk), while it is proportional
to the segment size $a$ of subsystem-A. Whereas
the Icebergs entropy includes contributions from rest of the subleading terms
 in small ${a\over b}$ expansion. But we immediately realize that 
all these terms in \eqn{fin2ab} are actually inseparable from each other. 
That is to say they are all equally important because the total sum of them  
depends only on single variable $l$ and nothing else. 
The total entropy of `bath plus system' 
thus has a constant value, with size $l$ being fixed.
Had we tried to ignore  subleading icebergs contributions in \eqn{fin2ab}, 
due to their smallness, 
we would find that total entropy $S[AUB]$  starts depending on $a$ or $b$
in a strange way. This might lead to wrong conclusions regarding the 
Page curve for bath entropy!
Hence we  conclude that $S_{Island}$ and $S_{Icebergs}$
should not be treated separate from leading bath entropy  in any situation. 
Furthermore, due to this the islands and icebergs will actually remain invisible, 
as these contribute
only in eq.\eqn{fin2} which represents an unphysical (higher value) 
extremum of bath 
entropy as per quantum entropy proposal \eqn{fin2s}.\footnote{The 
quantum entropy proposal might not work if the RT formula  in eq.\eqn{fin2} 
does not correctly represent the entropy of large CFT
bath system in a multi-system entanglement. I thank
S. Theisen for fruitful discussion over this point.  }
It will get clarified further as we proceed.

\subsection{\bf Generalization: Entropy of multiple quantum dots}

\begin{figure}[h]
\centerline{\epsfxsize=3in
\epsffile{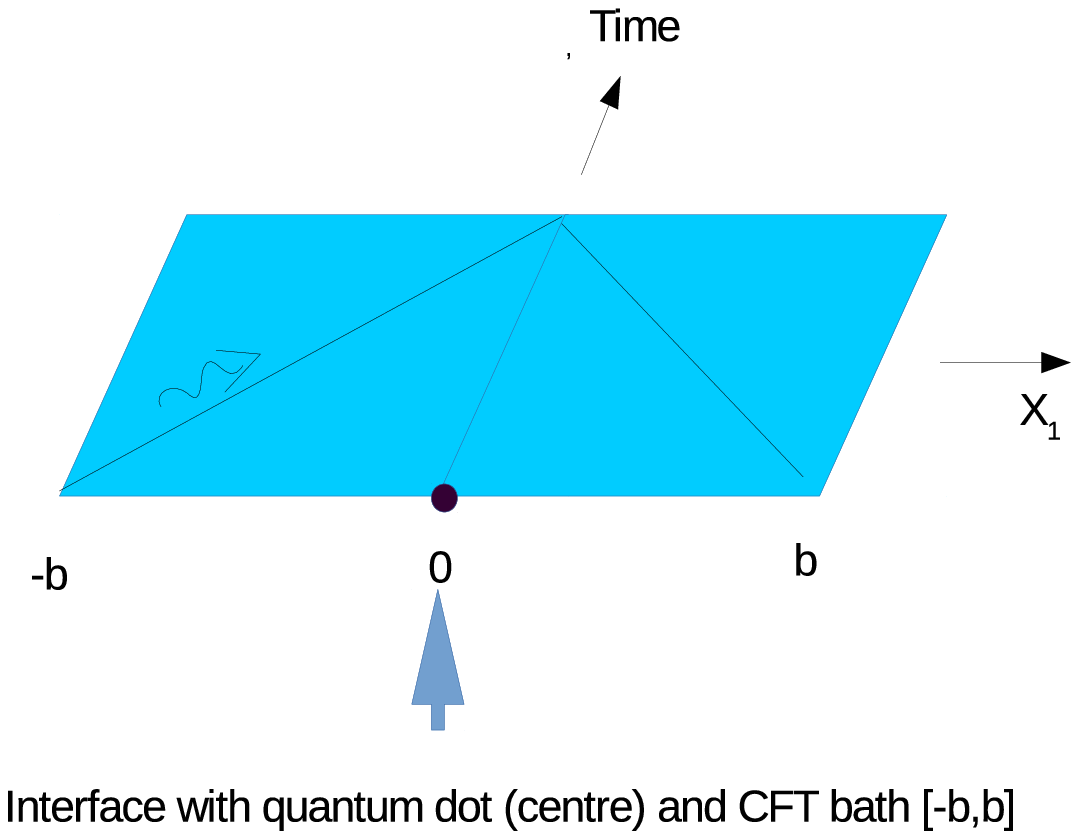} }
\caption{\label{fig21b} 
\it A typical symmetrical 
 arrangement of a quantum dot system (at the centre of $x$-axis) and 
1-dimensional bath of size $b$ on either side. The boundary CFT theory is described in 
2-dimensional Minkowski spacetime. }
\end{figure} 
 
We now discuss a limiting case of the entropy in  eq.\eqn{fin2}. 
 Considering  very small size limit $a\to R$, where $R$ is Kaluza-Klein
scale and $R \gg\epsilon$.
So  we can express $a= 2\pi n R$, with $n$ being discrete. 
Note we are simply assuming that there exists an
intermediate Kaluza-Klein scale $(2\pi R)$ at shorter distances.
That is the system-A appears essentially point or dot like. 
(So that it becomes  plausible to consider a 
dual JT gravity interpretation of dot-like system-A.)
 For brevity we shall only take small 
 $n$ values, also $R$ is taken really small 
(If there is any difficulty we will simply take $n=1$). The 
system-A now can be treated as a quantum-dot (point) sandwiched 
between finite CFT bath of size $b$ on both sides. We have the situation
resembling as in figure \eqn{fig21b}.  

The expansion
on the r.h.s of \eqn{fin2}, for  small ratio $({2\pi n R\over b})\ll 1$, gives us
\bea\label{fin2an}
S_{bath}[B]
&&= 
{L \over 2 G_3}\left( \ln{2b \over \epsilon} + {2\pi n R\over b}- 
{1\over 2} ({2\pi n R\over b})^2 +\cdots \right)
+{L \over 2 G_3}\ln{4\pi n R \over \epsilon}\br
&&\equiv 
\left(S^{(0)}_{bath} + S_{Island}
+ S_{icebergs}\right)
+{L \over 2 G_3}\ln{4\pi n R \over \epsilon}
\eea
where the expressions are identified as 
\bea
&&S_{Island}=
{L \over 2 G_2} {n\over b}, ~~~
\br &&S_{icebergs}=-{L \over 4 G_2} 
 ({2\pi R n^2 \over b^2}) + O(n^3) 
\eea
The $G_2$ is 2-dimensional Newton's constant.
The islandic contribution, especially for $n=1$, 
is precisely the gravitational entropy
of an island boundary (located at $z=b$), as first discussed by \cite{almheri}, while
the icebergs entropy includes rest all subleading contributions and there are 
infinitely many such terms in the series. We note that
the net contribution of all iceberg terms is however overall negative. 
In reality it will not be possible to separate them from 
first two leading terms in \eqn{fin2an} at all!
The icebergs become important because the total contribution 
of these terms (within the parenthesis) does add up to 
a constant, $l$ dependent quantity (as full size $l$ is fixed). 
It is quite clear from starting line of the 
perturbative expansion in \eqn{fin2an}. 
While the entropy of the q-dot in the center is
\bea
 S_{dot}={L \over 2 G_2}{1\over 2\pi R}\ln{4\pi R n\over \epsilon} 
\eea
Thus in conclusion, the island or icebergs individual 
contribution is of no physical significance,
as the Page-curve of bath CFT matter (of bath-B) will be fully determined by the entropy 
 of the system-A  (quantum KK dot) only.
It may be simply stated as
\bea
S_{bath}^{quantum}=_{min}\{S_{dot}, S_{bath}\}=
S_{dot}&
=&{L \over 2 G_2}{1\over 2\pi R}\ln{4\pi n R \over \epsilon}\br
& \propto &(
{L\over \pi R}\ln 2n + A_0)
\eea
where $A_0={L\over \pi R}\ln{2\pi R\over \epsilon}$ is a known area constant.
It is the net bath entropy near the end of the Page curve
and it entirely gets contribution
from the smallest RT surface corresponding to the KK q-dot. 
Note the dot is not exactly
a point instead it has size $\approx O(2\pi R)$.
Importantly the entropy comes as quantized  $\sim {\ln 2n}$, where $n=1,2,3,...$. 
Obviously we could trust our results for small $n$ only. For very large $n$ it would be
better to go for noncompact geometry. 
If this is true then the Page curve will necessarily show 
{\it discrete} jumps as and when $n$ value jumps. This 
can be taken as an example of
 strongly coupled system of $2n$ q-dots 
at the centre of a large
 CFT bath. These conclusions will not alter even if we take 
an infinite bath limit $(b\to\infty, ~l\to\infty)$. 
We conclude that we  should not see islands 
and icebergs individually,
as their total contributions leads to a constant value only.

\section{Entropy at finite temperature}
We now study the case of entropy
at finite temperature when boundary CFT is in a mixed state. 
Here the bulk  AdS geometry has a black hole horizon
\bea\label{btz23}
ds^2={L^2 \over z^2}(-f(z) dt^2 +{dz^2 \over f(z) } + dx^2)
\eea
The function $f(z)=(1-{z^2\over z_0^2})$ with
 $z=z_0$ is location of black hole horizon. So
there is a finite temperature in the boundary field theory.
\footnote{One may also take compact coordinate  $x=L\phi$, with 
range $0\le \phi \le 2\pi$ for the BTZ
black holes background.}  Now
the system-A of size $2a$
 is taken to be in thermal equilibrium with symmetrical bath,
of total size $2b$, and system-A is located in the middle 
of bath system-B of size $2b$. 

 We only discuss the case when $a\ll b$. The  entropy of the finite temperature
CFT bath system is then given by 
\bea\label{ft1}
S[B]= 
{L\over 2G_3} \ln \sinh({l\over z_0})  
+{L\over 2G_3} \ln \sinh({l-2b\over z_0})  +UV~ terms
\eea
The first term on the r.h.s. is just a system constant.
While the  entropy  for the system-A will be 
 \bea\label{fin1a}
S[A]=  
{L \over 2G_3}  \ln \sinh{l-2b\over z_0}+ S_{UV}.
\eea
Once again we can make out that the two expressions \eqn{ft1} and \eqn{fin1a}
above differ by an over all constant
term that depends upon the total size $(l)$. 
Actually  both these entropies are two extremas.
But the quantum entropy of the  bath subsystem  
would only come from
 the minimum of the two values
 \bea\label{fin17}
S_{quantum}^{b\gg a}[B]=_{min} \{S[A], S[B]\}=
{L \over 2G_3}  \ln \sinh{l-2b\over z_0}+ S_{UV}.
\eea

\begin{figure}[h]
\centerline{\epsfxsize=3.2in
\epsffile{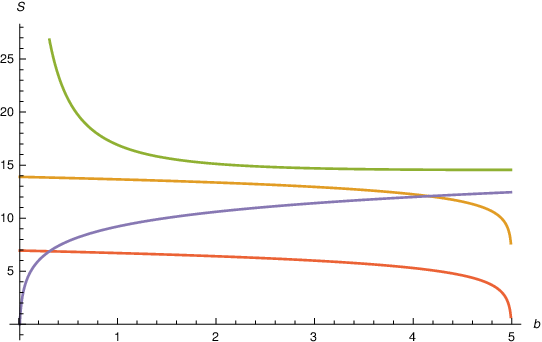} }
\caption{\label{fig211}
\small\it
Plots of 
$(2S_{UV} + Log \sinh (l/z_0) + Log \sinh(l - 2 b)/z_0)$,
$(2S_{UV} + Log \sinh (2b/z_0) + {2a\over z_0 \tanh(2b/z_0)})$,
$(S_{UV}+Log \sinh(l - 2 b)/z_0)$, and $(2S_{UV} + 2 Log \sinh (b/z_0))$
for parametric choice $l=10,~ z_0=20$, and $S_{UV}\simeq 7.6$. 
Only the lowermost falling graph (in red)
is good for entropy of large bath at finite temperature as 
per quantum entropy proposal.
The upper falling curve (which includes all island and icebergs entropies)
represents an unphysical extremum for bath entropy. 
The topmost curve (in green) which includes only islandic contribution is rather
a crude approximation and untrustworthy.
The sole rising graph (in blue) is good for entropy of small size $(b \sim 0)$ baths
only. 
We have  set $ {L\over 2 G_3}=1$.}
 \end{figure}

An interesting case arises  when $a\simeq  2\pi n R$, 
i.e. $b= {l\over 2}-2\pi n R$, 
we might wish to expand the r.h.s. of eq.\eqn{ft1} as
\bea\label{fin2am}
S[B]=
{L \over 2 G_3}  
\ln\sinh{l\over z_0} 
+{L \over 2 G_3}\ln\sinh{4\pi  n R \over z_0}+ UV~terms.
\eea
Note we may always express
\bea
{L \over 2 G_3}
\ln\sinh{l\over z_0} \equiv 
{L \over 2 G_3} \ln\sinh{2b \over z_0} 
+ S_{Island}+ S_{icebergs}=Fixed
\eea
where gravitational contribution from island boundary (at $z=b$) is
namely \bea
S_{Island}=
{L \over 2 G_3}  
{ 2a \over z_0\tanh{ 2b\over z_0}}=
{nL \over 2 G_2}  
{ 1 \over z_0\tanh{ 2b\over z_0}}=n \cdot S_{n=1} ~~~
\eea
which is like  entropy of $n$ independent islands. While other
 subleading (icebergs) terms in the series
are  important 
for generalized entropy, as altogether these add up to give just a constant 
 entropy, $ {L \over 2 G_3}  
\ln\sinh{l\over z_0}$. If we ignored any one of them under any assumption 
we cannot arrive at this conclusion. 

For the central dot-like system-A 
(which may also be thought of  as dual of JT gravity) the entanglement entropy is
\bea
S_{dot}={1 \over 2 G_2}{L\over 2\pi R}\ln\sinh{4\pi n R \over z_0}+  S_{UV}
 \eea
which is the smallest entropy value and it is also quantized, by virtue of
 KK scale. However we should trust this result for small and discrete $n$ values
only.

Note that in infinite bath limit, the geometric entropy  of the island boundary
tends to becomes the horizon entropy
(for ${b\to \infty}$ )
\bea
S_{island} \to   {n L \over 2 G_2 z_0}\equiv 2n \cdot S_{BH}
\eea
But we should not isolate it from the entropy  of the icebergs, 
because in totality the
island and icebergs entropies along with the leading `pure' bath entropy
term give overall constant quantity,  
depending only on the size $l$. Note length $l$ also gives a measure
of total energy of the bath and the subsystem, so it has to be conserved 
unless there is a leakage from the bath to the outside. We have not considered 
that hypothesis here or other time evolving cases. It is possible that $n$ can jump
in a time dependent process. The discrete nature of the bath entropy is a consequence of
the Kaluza-Klein scale at small distances. If there is no such compactification
at short distances then the bath entropy will vanish smoothly when $a\to 0$.

Thus from \eqn{fin17} for large CFT bath (in contact with quantum dot system)
the quantum entropy 
at finite temperature can simply be written as
\bea
S_{bath} \equiv S_{dot}
={1 \over 2 G_2}{L\over 2\pi R}\ln\sinh{4\pi n R \over z_0}+  S_{UV}.
\eea
It is smallest of the entropies and  has no contribution 
from an island or the icebergs. It also implies that the island and icebergs 
cannot be independently observed. 
They would remain fictitious because they contribute only 
to an unphysical higher entropy branch arising from  eq.\eqn{fin2am}.

\section{Summary}
We have proposed that quantum entropy of entanglement of 
a large CFT bath system follows the minimality principle that
$$ 
S_{bath}= {}_{min} \{S_{dot}, S_{dot}+Const. \}
={L\over 4\pi G_2 R}\ln\sinh{4\pi n R \over z_0}+  S_{UV}.
$$
and thus it realizes the Page curve for the entropy of thermal CFT matter
in the bath. This conclusion is based upon the
observation that q-dot and the large bath entropies differ only by an overall constant.
The constant part of entropy depends only on total system size $(l)$, and that remains 
fixed following the conservation laws.  
 
We have found that islands and the icebergs only contribute to the
unphysical extrema of higher entanglement entropy. 
The actual physical extremum with lower
entropy however never gets contribution from 
these fictitious elements. Therefore the entropy formula such as \eqn{ficti1}
may not give complete picture of entanglement
as it ignores subleading contributions, such as the iceberg terms. 
In fact there is an infinite series of them. We also 
find that our results are quite generic and may be extended to
higher dimensional systems also. These would be reported separately.


\vskip.5cm
   


\end{document}